%% file: main2.tex
\documentclass[conference]{IEEEtran}

\usepackage{svg}
\usepackage{enumitem}
\usepackage{longtable,booktabs}
\usepackage{array}
\IfFileExists{footnotehyper.sty}{\usepackage{footnotehyper}}{\usepackage{footnote}}
\makesavenoteenv{longtable}

\usepackage{multirow}
\usepackage{tablefootnote}
\usepackage{colortbl}
\usepackage{graphicx}
\usepackage{acro}
\usepackage{siunitx}
\usepackage{tablefootnote}

\usepackage{bbding}
\usepackage{fontawesome} 
\usepackage{transparent}
\usepackage{soul}
\usepackage{flushend}
\setul{0.5ex}{0.3ex}  %
\setulcolor{red}
\definecolor{lightgray}{gray}{0.925}

\input{acronyms}
\begin{document}

\title{On the Resilience of 5G NR Against Jamming}

\author{
\IEEEauthorblockN{
Sotiris Michaelides\IEEEauthorrefmark{1},
Fidelius Lula\IEEEauthorrefmark{1},
Daniel Eguiguren Chavez\IEEEauthorrefmark{1},
Michael Rademacher\IEEEauthorrefmark{2}\IEEEauthorrefmark{3},
and Martin Henze\IEEEauthorrefmark{1}\IEEEauthorrefmark{2}
}
\IEEEauthorblockA{
\IEEEauthorrefmark{1}\textit{RWTH Aachen University},
Aachen, Germany, 
\IEEEauthorrefmark{2}\textit{Fraunhofer FKIE},
Bonn/Wachtberg, Germany\\
\IEEEauthorrefmark{3}\textit{University of Applied Sciences Bonn-Rhein-Sieg},
Sankt Augustin, Germany
}
\IEEEauthorblockA{
\{michaelides, henze\}@spice.rwth-aachen.de,
\{fidelius.lula, daniel.eguiguren\}@rwth-aachen.de,\\
michael.rademacher@fkie.fraunhofer.de
}
}

\maketitle

\begin{abstract}
  With the increasing use of 5G networks in availability-critical systems, including industrial networks and critical infrastructure, a comprehensive understanding of their resilience to cellular jamming has become imperative.
  However, research so far has focused on isolated evaluations under fixed 5G physical-layer configurations, making it difficult to perform sound comparisons, for example, to identify differences between frequency bands or channel bandwidths w.r.t. jamming resilience.
  To fill this gap, we implement and release an open-source 5G jammer within the ns-3 simulator.
  Leveraging the controllable determinism of simulation, we compare the impact of cellular jamming across different dimensions such as cellular technologies, frequencies, subcarrier spacing, and channel bandwidth.
  Our results show that while cellular technology and subcarrier spacing have little impact, channel bandwidth and frequency range play a critical role in improving jamming resilience.
  \end{abstract}
  \IEEEpeerreviewmaketitle


  \section{Introduction}
  \label{sec:intro}
  
  Jamming attacks have posed a substantial threat since the early days of mobile communications and remain a persistent challenge today~\cite{jammingThreatAss,controljamm}. 
  These attacks disrupt legitimate communication by transmitting signals on the same frequency as their target, thereby causing interference~\cite{jamming4G5G,birutis2022study5G}. 
  Despite extensive research and mitigation techniques, complete immunity to jamming is practically infeasible unless both transmitter and receiver are fully isolated~\cite{michaelides2025industry5G}.

  The criticality of jamming has intensified with the advent of fifth-generation (5G) mobile networks, which, unlike their predecessors, are increasingly deployed in industrial and mission-critical environments~\cite{michaelides2025industry5G}. 
  Such networks prioritize operational availability, as service disruptions can have severe societal and economic consequences and may even endanger lives~\cite{michaelides2026ransecurity}. 
  In this context, even barrage jamming—where an adversary continuously transmits high-power signals~\cite{JammingSurvey}—can compromise availability or violate the strict latency requirements~\cite{michaelides2025industry5G,michaelidesLatency2025}.
  A comprehensive understanding of 5G’s robustness to such attacks is therefore essential.
  
  In particular, 5G New Radio (NR) introduces transformative physical layer (PHY) characteristics—such as scalable \ac{scs}, channel bandwidths, and mmWave operation. 
  While these features provide unprecedented flexibility and performance gains, they also create a large configuration space that may fundamentally alter a network’s susceptibility to jamming~\cite{birutis2022study5G,magnetism5020010}. 
  Consequently, systematic evaluation is essential to identify which features enhance or reduce robustness, providing actionable guidance for mission-critical deployments where availability is paramount~\cite{michaelides2025industry5G}.

  While prior work has attempted to evaluate 5G’s resilience to jamming~\cite{BIRUTIS202258,magnetism5020010,controljamm,jammingfr1}, the flexibility of the 5G PHY and practical constraints often limit these studies to isolated, single-PHY configurations (cf.~\S\ref{sec:related}). 
  This restricts the generalizability of their findings and fails to capture the nuances of 5G’s massive parameter space, leaving a critical gap in our understanding. 
  Consequently, there is a need for a controlled, reproducible, and systematic evaluation framework capable of assessing jamming resilience across the full breadth of 5G PHY configurations.
  
  However, deploying such a framework in the real world is highly challenging. (i) Spectrum license acquisition is complicated, costly, and rarely covers the full spectrum, (ii) 5G networks are expensive, limiting experiment replicability, and (iii) jamming devices are illegal to acquire or operate in many countries~\cite{fccJammersEnforcement,JammerMasterGermany}. Moreover, (iv) fluctuations in physical channel conditions make it impossible to compare different PHY configurations under identical conditions.

   \noindent\textbf{Contributions.}
  In this paper, we address these challenges by adding jamming capabilities to a proven and widely used network simulator with detailed 5G support. 
  Leveraging the controllable determinism of network simulation, we perform sound comparisons across diverse 5G PHY dimensions -- including frequency ranges (sub \SI{6}{\giga\hertz} vs. mmWave), subcarrier spacings (\SI{15}{} vs. \SI{30}{} vs. \SI{60}{} vs. \SI{120}{\kilo\hertz}), and channel bandwidths (\SI{20}{} vs. \SI{50}{} vs. \SI{400}{\mega\hertz}) -- evaluating their resilience to cellular jamming and directly comparing them with 4G LTE.
  More concretely, our contributions are:
  
  \begin{enumerate}[topsep=0pt,leftmargin=14pt]
    \item We identify challenges in researching cellular jamming, especially w.r.t.\ sound comparisons such as identifying differences across frequency bands (§\ref{sec:background}).
    \item We present the first open-source 5G jammer implemented in the ns-3 simulator, enabling \emph{safe, legal, and cost-free} jamming experimentation across all 5G frequency bands, incl. mmWave~(§\ref{sec:simulation}).
    \item We assess several 5G physical-layer parameters and their impact on jamming resilience, including a practical comparison with 4G LTE (§\ref{sec:resilience}), to quantify robustness differences that have previously only been analyzed theoretically~\cite{birutis2022study5G,jammingThreatAss} and to provide guidance for jamming-resilient network deployments~(§\ref{sec:implecations}).
  \end{enumerate}
  
  \noindent\textbf{Availability Statement.}
  To spur further research, we open-source our jammer implementation~\cite{ns3nrjammer2025}.

  \section{Jamming in 5G Networks}
  \label{sec:background}
  The 5G PHY provides unprecedented flexibility to support a wide variety of deployments. 
  In this section, we highlight its key changes from 4G LTE to 5G NR (\S\ref{sec:phy}) and briefly explain the concept of jamming (\S\ref{sec:jamming}). 
  We then outline the main challenges of experimental 5G jamming research (\S\ref{sec:challenges}) and review related work to illustrate these challenges in real-world experiments (\S\ref{sec:related}).
  
  \subsection{5G PHY in a Nutshell}
  \label{sec:phy}
  At a high level, 5G retains the same architectural components as 4G LTE -- namely the \ac{ue}, the \ac{ran}, and the core network~\cite{michaelides2025industry5G}. 
  However, 5G New Radio (NR) introduces substantial PHY enhancements to support higher data rates, lower latency, and diverse deployment scenarios, raising the question of how the 5G PHY resists jamming attacks.
  
  The 5G PHY, operating between \acp{ue} and the \ac{ran}, encodes data for transmission, maps it to radio waveforms, and decodes received signals. 
  A defining feature of 5G is its operation across a broader spectrum. 
  It supports sub-\SI{6}{\giga\hertz} frequencies in Frequency Range 1 (FR1) and millimeter-wave frequencies above \SI{24}{\giga\hertz} in Frequency Range 2 (FR2)~\cite{jammingfr1}. 
  FR1 enables channel bandwidths up to \SI{100}{\mega\hertz}, while FR2 supports up to \SI{400}{\mega\hertz}; in contrast, LTE is limited to \SI{20}{\mega\hertz} per carrier~\cite{birutis2022study5G}.
  
  To efficiently utilize these wide bandwidths, 5G NR (like LTE) employs an OFDM-based air interface. 
  Data is transmitted in parallel over orthogonal subcarriers, each carrying OFDM symbols over short, fixed intervals. 
  Unlike LTE, which uses a fixed 15 kHz \ac{scs}, 5G introduces scalable numerology with SCS values ranging from 15 to 240 kHz~\cite{birutis2022study5G,smart}. 
  Larger \ac{scs} values reduce symbol duration and enable lower latency operation~\cite{5Glatency}. 
  The combination of subcarriers and OFDM symbols in time forms a two-dimensional time–frequency resource grid, which constitutes the fundamental transmission structure of the 5G air interface~\cite{jammingThreatAss}.
  
  Within this grid, 5G dynamically maps data, control, and synchronization signals onto time–frequency resources. Physical channels carry encoded information, while physical signals provide reference waveforms for synchronization, channel estimation, and measurements~\cite{jammingThreatAss}. 
  Additionally, 5G extends LTE’s MIMO capabilities by supporting larger antenna arrays and directional beamforming, enabling spatial multiplexing and improved coverage and capacity~\cite{birutis2022study5G,smart}. 
  Despite these enhancements, 5G NR—like all wireless systems—remains vulnerable to intentional interference, a.k.a. jamming, which we explain in the following.

  \begin{figure}[t] 
    \centering
    \includegraphics[width=0.65\columnwidth]{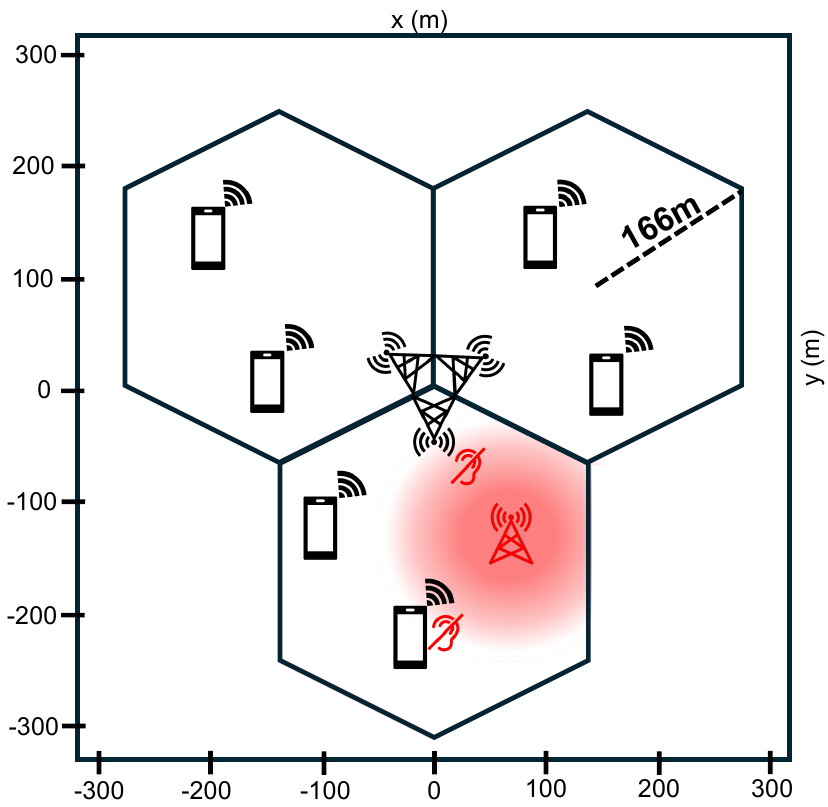} 
    \caption{Example topology with six UEs, one \textcolor{red}{barrage jammer}, and three base stations located at the center. Each base station serves its respective cell using a distinct portion of the frequency band to reduce interference.}
  
    \label{fig:jammer}
  \end{figure}
  
  \subsection{Jamming Attacks}
  \label{sec:jamming}
  Jamming attacks aim to disrupt communication between \acp{ue} and the \ac{ran} by transmitting interfering signals within the same frequency band at the same time~\cite{JammingSurvey}.
  Jamming strategies are typically categorized as \emph{barrage jamming} or \emph{smart jamming}~\cite{jamming4G5G}. 
  Barrage jamming emits high-power noise continuously, across a wide frequency range, degrading the \ac{snr}, thereby effectively deafening receivers to legitimate signals (cf. Fig.\,\ref{fig:jammer}). 
  Such attacks are often used as a baseline for evaluating smart jamming attacks~\cite{JammingSurvey}, which selectively target protocol channels, control signals, or procedures. 
  Rather than continuously transmitting noise, the attacker focuses interference on critical resources to maximize disruption while minimizing power consumption. 
  By exploiting protocol-specific vulnerabilities, smart jamming is stealthier; however, it requires substantial effort, synchronization, and detailed system knowledge to execute effectively.
  In this paper, we focus on barrage jamming, as it is a far more likely attack scenario due to its lower effort and serves as a widely accepted baseline. 
  We also rely on simulation, since real-world experiments are technically, financially, and regulatorily challenging.

  \subsection{Challenges in Jamming Research}
  \label{sec:challenges}
  Practical jamming research remains limited due to several constraints.
  First, lawful experiments require spectrum licenses, which are expensive and typically restricted to specific frequency bands, from national regulatory authorities.
  Further, in some jurisdictions intentional jamming is prohibited altogether—even for research purposes~\cite{fccJammersEnforcement,JammerMasterGermany}.
  Second, deploying a commercial-grade 5G PHY testbed is costly~\cite{5Gcube,michaelides2025industry5G}. Consequently, many studies rely on open-source frameworks such as OAI and srsRAN, which may not support the full range of PHY configurations, particularly w.r.t.  channel bandwidth and \ac{scs}~\cite{OpenAirInterfaceRAN,srsranConfigRef}.
  Finally, to the best of our knowledge, no publicly available mmWave jammer suitable for controlled research exists, severely limiting experimental studies in FR2. 
  Even if such a jammer were available, experiments at higher frequencies—such as the 5G mmWave range—would remain challenging. 
  Millimeter-wave channels are significantly more dynamic and sensitive to blockage and environmental variations compared to sub-6\,GHz systems, making it difficult to reproduce identical channel conditions across experiments~\cite{6834753}.
  Lastly, beyond these technical constraints, additional ethical considerations further complicate jamming research. For instance, jamming signals may unintentionally interfere with and disrupt critical services, such as emergency communication systems.
  Next, we review related work in 5G jamming research, highlighting the challenges and constraints associated with conducting such studies in real-world deployments.

  \subsection{Related Work and the Benefits of Simulation}
  \label{sec:related}
  Existing experimental studies consistently demonstrate the susceptibility of 5G NR to both wideband and targeted interference. 
  For example, measurements in a commercial network showed that service becomes intolerable when barrage jamming exceeds the signal by approximately 5 dB~\cite{BIRUTIS202258}. 
  However, the experiments were limited to band n78 with \SI{30}{\kilo\hertz} SCS and \SI{80}{\mega\hertz} bandwidth. 
  Similarly, Varga et al. demonstrated that wideband jamming can fully disrupt communication, yet their experiments were confined to band n78 with \SI{40}{\mega\hertz} bandwidth~\cite{jammingfr1}. 
  Flores et al. achieved zero throughput for specific UEs using targeted jamming, but again under a single configuration (band n78; \SI{30}{\kilo\hertz} SCS; \SI{40}{\mega\hertz} bandwidth)~\cite{SmartJamming}.
  Likewise, Mahmudul et al. performed targeted jamming attacks on the control channels of 5G NR (band n71; \SI{657.5}{\mega\hertz}) and found that their approach could successfully induce denial of service conditions at lower jamming power levels compared to conventional methods~\cite{controljamm}.
  Beyond intentional interference, Saleem et al. examined transient electromagnetic interference to compare LTE-A and 5G NR resilience~\cite{magnetism5020010}. 
  Although LTE showed superior robustness, the study was restricted to the \SI{2.2}{\giga\hertz} band.
  
  Overall, existing work confirms the practical vulnerability of 5G systems to jamming while highlighting the difficulty of conducting comprehensive experiments. Most studies focus on a single PHY configuration, limiting generalizability and to other configurations—particularly for mmWave deployments, where experimental results are largely absent. 
  To address this gap, we turn to simulation, which provides a safe, cost-effective environment for testing scenarios that are currently challenging in real-world deployments (e.g., mmWave jamming). 
  Simulation also allows easy reconfiguration of the network while keeping channel conditions identical, ensuring observed effects stem directly from PHY changes. Finally, it enables us to open-source our jammer~\cite{ns3nrjammer2025}, whose implementation we discuss next, paving the way for future works.

  \section{Simulating Cellular Jamming} 
  \label{sec:simulation}
  
  The Network Simulator 3 (ns-3) is a widely recognized open-source discrete-event simulation platform, extensively used in academia for cellular network research~\cite{5glena_papers_website}. 
  Here, we provide an overview of ns-3 and the \texttt{5G-LENA} module (\S\ref{sec:ns-3}), detailing their capabilities for realistic 5G simulation, before describing our jammer implementation within this framework (\S\ref{sec:jammerimpl}), which was developed using ns-3 v3.39 and \texttt{5G-LENA} v2.5.y.

  \subsection{Basics of ns-3 and 5G-LENA}
  \label{sec:ns-3}
  An ns-3 simulation is structured around nodes, serving as containers for network entities such as UEs. 
  These nodes house \texttt{NetDevices} (network interfaces) that connect to \texttt{Channels}, representing the communication medium and its propagation characteristics. 
  Each node is configured with a protocol stack and applications to generate traffic, while helper classes are used to automate tasks such as setting up network addresses.
  The simulator operates as a discrete-event engine, advancing time between scheduled events (e.g., transmissions), and provides tracing mechanisms for post-simulation analysis of metrics such as throughput and latency.
  For advanced wireless simulations, ns-3 uses the \texttt{Spectrum} module~\cite{spectrumModule}, which models signals in the frequency domain and enables detailed interference and propagation analysis~\cite{RademacherInterferenceNs3}. 
  Its core abstractions, \texttt{SpectrumChannel} and \texttt{SpectrumPhy}, represent the communication medium and the physical-layer entity, respectively. 
  Received \ac{psd} values account for antenna gains, propagation loss, and channel effects. 
  Interference is aggregated in the \texttt{MultiModelSpectrumChannel}, as additive Gaussian noise to compute the~\ac{snr}~\cite{5glena_nr_manual_bal2009}.
  
  In our study, we model 5G NR networks using the 5G-LENA extension~\cite{5gLENAfeatures}, built upon the \texttt{mmWave} module~\cite{PATRICIELLO2019101933,mmwavemodule}. 
  It provides 5G NR PHY/MAC layer  support  to the existing (LTE) LENA module. 
  Crucially for jamming studies, 5G-LENA implements key 5G PHY features such as scalable bandwidths and \ac{scs}, beamforming, and support for both FR1 and FR2~\cite{5gLENAfeatures}.
  The framework is calibrated against 3GPP TR 38.901 channel models~\cite{3GPP_TR_138_901_V19_2_0}, ensuring that propagation characteristics, interference, and performance match those observed in real-world 5G networks~\cite{calibration}. 
  This calibration increases confidence that the simulation results are accurate and reproducible, despite the inherent abstractions and simplifications of the simulation model.
  Such high-fidelity modeling makes ns-3 a suitable, safe, and cost-effective platform for evaluating jamming impact.

  \subsection{Jammer Implementation}
  \label{sec:jammerimpl}
  
  At the core of our implementation is the \texttt{WaveformGenerator} from the ns-3 Spectrum module, which models a simple device emitting power into the spectrum to generate interference~\cite{ns3_waveform_generator}. 
  In its original form, the \texttt{WaveformGenerator} inherits from the abstract \texttt{SpectrumPhy} class and generates signals based on spectrum signal parameters, including \ac{psd}, bandwidth, and duration, which are then propagated through ns-3 channel models. 
  Effectively, it defines the jamming power per bandwidth unit (cf.~\S\ref{sec:wideband}), along with timing,  determining how the signal propagates and interferes with other transmissions.
  To effectively disrupt communication in 5G environments, we extend this functionality by implementing the \texttt{NrJammingGenerator} class, which inherits from NrSpectrumPhy as provided by the 5G-LENA framework. 
  This design allows us to model interference based on the physical layer characteristics defined by the 5G NR standard (e.g., SCS, bandwidth, and beamforming), rather than treating it as technology-agnostic noise.
  Our jammer effectively simulates a malfunctioning/compromised or malicious NR device (e.g., a base station). 
  By utilizing the calibrated classes of 5G-LENA without modification~(cf.~\S\ref{sec:ns-3}), we ensure their effectiveness and accuracy in reflecting real-world conditions, allowing our jammer to generate interference based on realistic channel characteristics. 
  To streamline the configuration and deployment of our jammer, we provide a helper class that simplifies setup by handling tasks such as channel assignment, bandwidth configuration, and PSD specification. 
  Our high-fidelity, easily configurable jammer enables direct evaluation of its impact across different 5G NR PHY configurations, as described in the following section.

  \section{Methodology}
  By deploying our jammer within an existing network scenario integrated into 5G-LENA (\S\ref{sec:scenario}), we conduct a series of jamming experiments under different 5G and LTE PHY configurations and monitor network performance metrics to assess the impact (\S\ref{sec:measurements}). %
  
  \subsection{Deployment Scenario}
  \label{sec:scenario}
  
  We use the \texttt{lena-lte-comparison-user} scenario from the 5G-LENA module, designed for 4G LTE and 5G NR comparisons~\cite{5glena_nr_manual_bal2009}. 
  The scenario models a multi-cell hexagonal grid with sectorized sites. 
  In our experiments, we use three sectors to reduce simulation time, as shown in Fig.~\ref{fig:jammer}, which provides an exemplary deployment with one jammer. 
  Each sector operates on a separate portion of the frequency band, as in real deployments, enabling spectrum reuse~\cite{5glena_nr_manual_bal2009}. 
  This aligns with our goals, as it eliminates inter-cell interference and isolates the jammer’s effect.
  The scenario also supports 3GPP-calibrated channel models for micro, macro, and rural macro deployments and can emulate both LTE (via 5G-LENA/LENA) and 5G NR networks~\cite{calibration}. For LTE experiments, we use the 5G-LENA LTE implementation, which supports TDD multiplexing for fair comparison with FR2 bands (FR2 only supports TDD). 
  
  In our deployment (cf. Fig.~\ref{fig:jammer}), we use the micro-cell configuration. 
  Each sector has a radius of 166\,m and is served by a base station transmitting at 43\,dBm, serving two randomly distributed UEs. 
  Each UE receives a dedicated downlink UDP flow (20\,Mbps) from an external server. With the network deployment defined, we configure the 5G NR PHY parameters and monitor key performance metrics of the UDP flows to assess the jammer’s impact, as described next.

  \subsection{PHY Configurations and Measurements}
  \label{sec:measurements}
  
  In this study, we evaluate how SCS numerology, channel bandwidth, and operation in FR2 affect 5G jamming resilience. 
  Our PHY configurations are summarized in the upper part of Table~\ref{tab:phy-exp}, and only standard-compliant, technically meaningful configurations are considered.
  For instance, LTE is limited to 15 kHz SCS with a fixed channel bandwidth of 20 MHz in our setup, whereas wider bandwidths (e.g., 400 MHz) and higher SCS values (e.g., 120 kHz) are evaluated exclusively for NR FR2 (cf.~\S\ref{sec:phy}). 
  Each PHY configuration is evaluated across different combinations of experimental parameters, as summarized in the lower part of Table~\ref{tab:phy-exp}.
  
\begin{figure}[t] 
    \centering
    \includegraphics[width=0.92\columnwidth]{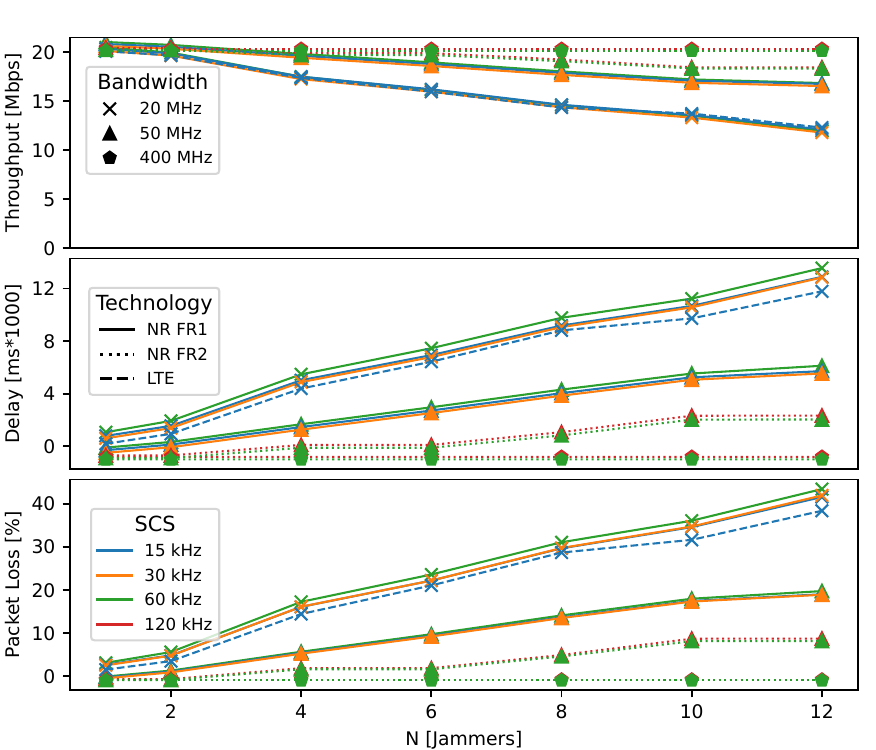} 
    \caption{Average flow performance at the lowest jamming power evaluated (15\,dBm) as the number of jammers increases. FR2 configurations consistently outperform others, with the 400\,MHz channel bandwidth largely unaffected.} 
  
    \label{fig:results}
  \end{figure}

  For each experiment, we randomly deploy jammers in cells and monitor network performance metrics to evaluate their impact. 
  A full simulation lasts 2 minutes, with jammers configured to interfere only with the corresponding sub-frequency of their assigned cell and becoming active 1 minute into the simulation, allowing the network to operate normally during the initial period. 
  Metrics are collected separately for each UDP flow and, at the end of the simulation, aggregated to compute the average performance. We focus on three key performance metrics: packet loss, flow latency, and throughput.
  These metrics are automatically collected by ns-3; however, we adjust the latency calculation to account for dropped packets, which are ignored in the default implementation. 
  Specifically, we include a penalty for lost packets by adding a fixed value equal to half the jammer’s active duration, reflecting the assumption that lost packets are approximately uniformly distributed over the jammer’s active period. 
  This adjustment ensures that latency metrics more accurately reflect the combined effects of delays and losses.
  Consequently, packet loss and latency results show similar trends (cf. Figs.~\ref{fig:results} \& \ref{fig:results2}).
  Each experiment is repeated 10 times with different initialization seeds. However, the same set of seeds and jammer placements is reused across all experiments, so that performance differences can be attributed solely to the PHY parameters.

  \begin{table}[t]
    \footnotesize
    \centering
    \renewcommand{\arraystretch}{1.2} %
    \caption{PHY Configurations and experimental dimensions}
    \label{tab:phy-exp}
    \begin{tabular}{|l|l|}
    \hline
    \multicolumn{2}{|c|}{\textbf{PHY Configurations}} \\ \hline
    Technology & 4G LTE, 5G NR \\ \hline
    Frequency Range (GHz) & FR1 (2.3), FR2 (26)    \\ \hline
    Subcarrier Spacing (kHz) & 15, 30, 60, 120  \\ \hline
    Channel Bandwidth (MHz) & 20, 50, 400  \\ \hline
    \multicolumn{2}{|c|}{\textbf{Experimental Dimensions}} \\ \hline
    Total Number of Jammers & 1, 2, 4, 6, 12 \\ \hline
    Jammer Power (dBm) &  15, 30, 45, 60, 75 \\ \hline
    \end{tabular}
    \end{table}

  \section{Analyzing Resilience to Jamming} 
  \label{sec:resilience}
  
  To highlight the impact of PHY configuration on jamming, we present results from both experimental dimensions (\S\ref{sec:results}): first, by increasing the number of jammers at the lowest jamming power to examine mild interference, and second, by increasing the jamming power with the maximum number of jammers to explore extreme conditions.
  We later identify and discuss the features that improve network performance under these jamming conditions~(\S\ref{sec:takeaway}).

  \subsection{Results}
  \label{sec:results}
  
  First, we consider jamming at the lowest tested power (15 dBm) while increasing the number of jammers.
  As shown in Fig.~\ref{fig:results}, clear performance groupings emerge.
  The first group includes all 5G FR1 configurations with 20 MHz channel bandwidth, along with LTE (15~kHz SCS, 20~MHz bandwidth), which perform very similarly. 
  With 12 jammers deployed, packet loss reaches roughly 45\%, corresponding to a throughput degradation of about 9–10 Mbps. 
  Interestingly, LTE slightly outperforms the other configurations in this group, consistent with prior work~\cite{magnetism5020010}, providing additional confidence in our results.
  The second group consists of FR1 configurations with 50 MHz channel bandwidth. These configurations perform better than the first group: at 12 jammers, packet loss is around 20\% and throughput degradation is approximately 4 Mbps.
  The third group includes FR2 configurations with 50 MHz channel bandwidth. These configurations show improved performance: with 12 jammers, packet loss reaches approximately 10\% and throughput degradation is around 2 Mbps. Flow latency remains below 10~ms for up to 6 jammers, indicating better performance than FR1 configurations but not as high as the wider FR2 channels.
  The final group comprises FR2 configurations with 400 MHz channel bandwidth. 
  These configurations maintain the highest performance under jamming: even with 12 jammers, packet loss remains at 0\%, full bandwidth is preserved, and transmission delays stay below 10 ms. 
  These results highlight that FR2 and wider channels can sustain higher throughput and lower latency under interference, illustrating the benefits of larger bandwidths for maintaining performance.

\begin{figure}[t] 
    \centering
    \includegraphics[width=0.9\columnwidth]{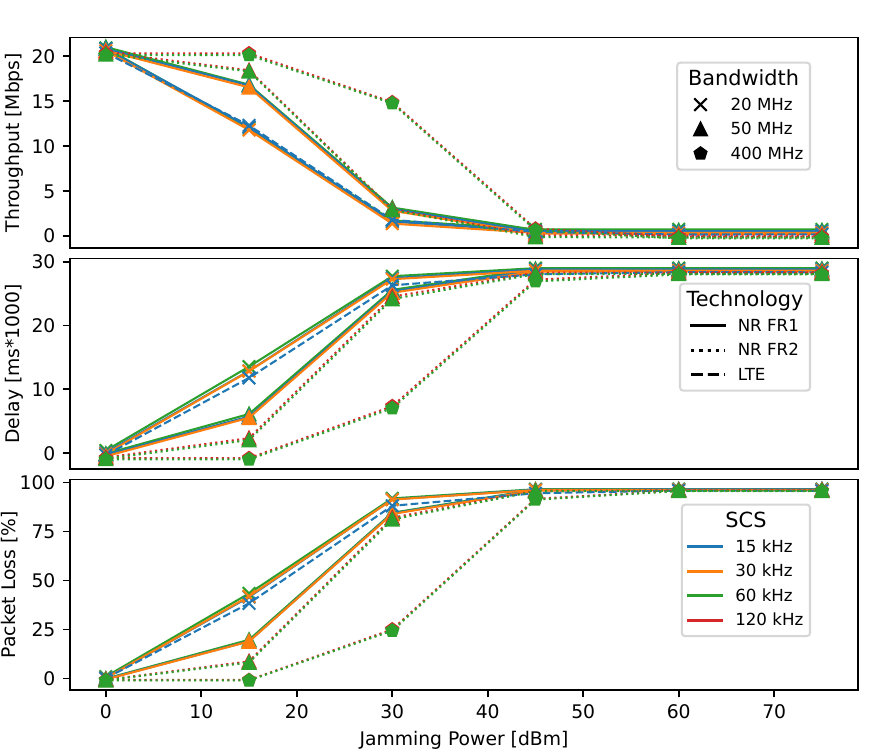} 
    \caption{Average flow performance with the maximum number of jammers evaluated (12) as their power increases. All configurations reach 100\% packet loss at 30\,dBm, except FR2 with 400\,MHz bandwidth which fails at 45\,dBm.} 
  
    \label{fig:results2}
  \end{figure}

  We next consider our second scenario, where 12 jammers are deployed, and the transmission power is increased to push the network to its limits (cf. Fig.~\ref{fig:results2}). 
  The same configuration groupings observed previously reappear. At 15 dBm of jamming power, the results are identical to those in the previous figure, so we focus on higher power levels.
  At 30 dBm, all configurations except FR2 with 400 MHz become unusable, with packet loss approaching 100\% and throughput near 0 Mbps. 
  FR2 with 400 MHz remains partially operational, maintaining roughly 15 Mbps throughput with around 20\% packet loss, but it also collapses at 45 dBm.
  These results further highlight the robustness of wideband mmWave against jamming. The next subsection examines the performance groupings and key takeaways of our study.

  \subsection{Key Takeaways and Discussion}
  \label{sec:takeaway}
  Our results reveal four clear performance clusters, showing that jamming resilience is driven primarily by \emph{frequency band and channel bandwidth}, rather than generation or SCS. 
  
  Based on our findings, we observe that:
  (i) LTE and 5G NR FR1 configured in the same frequency band and channel bandwidth exhibit near-identical degradation, regardless of the 5G SCS, indicating that 5G NR provides no inherent physical-layer advantage over LTE in this setting; and
  (ii) varying the SCS within either FR1 or FR2 has a negligible impact on packet loss or throughput, suggesting that SCS is not a decisive factor in anti-jamming robustness.
  
  However, (iii) operating in mmWave improves resilience compared to sub-6\,GHz bands, due to higher path loss and spatial confinement that also constrain the jammer (cf.~\S\ref{sec:spatial}); and
  (iv) increasing channel bandwidth plays the most significant role overall, as wider bandwidths spread jammer power across the spectrum and substantially increase channel capacity (cf.~\S\ref{sec:wideband}). 
  These effects are explained in detail in the following subsections.
  
  \subsubsection{The Spatial Dynamics of mmWave}
  \label{sec:spatial}
  Although mmWave signals experience higher propagation loss and are often considered more interference-prone, these same limitations also affect the jammer. 
  Faster attenuation reduces cell size~\cite{birutis2022study5G}, confining both legitimate and malicious transmissions. 
  As a result, a jammer must either transmit at higher power or move closer to its target to remain effective. 
  Jamming impact therefore depends on the relative power received by the legitimate signal compared to the jammer. %
  
\begin{figure}[t] 
    \centering
    \includegraphics[width=\columnwidth]{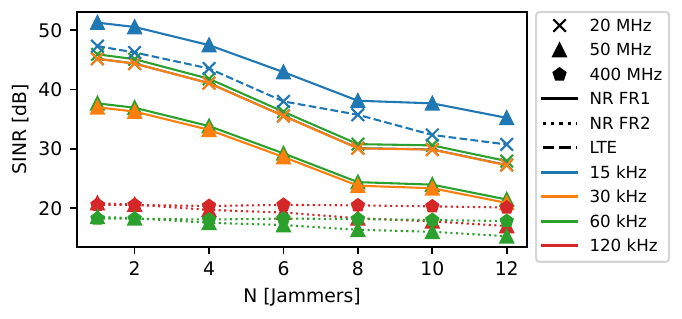} 
    \caption{Average SINR across all bands shows that sub-6 GHz configurations initially range between approximately 40-50 dB (considered excellent), while mmWave bands start around 20 dB (considered good). When 12 jammers are introduced, sub-6 GHz experiences significant SINR degradation, whereas mmWave is only slightly affected, highlighting the stronger spatial isolation characteristics of mmWave propagation.} 
  
    \label{fig:SINR}
  \end{figure}

  To verify this, we examined the simulation logs for our first scenario with constant 15 dBm jamming power to extract and present the \ac{snr} for each configuration in Fig.\,\ref{fig:SINR}.
  mmWave configurations operate with a baseline \ac{snr} of roughly 20~dB—lower than the 38–52~dB observed in sub-6\,GHz bands—yet degradation remains below 5~dB even with 12 jammers. 
  In contrast, sub-6\,GHz frequencies experience approximately 15~dB degradation under the same conditions. 
  This indicates that, in the mmWave band, the jamming signal was not strong enough to substantially impact performance.
  Overall, these results show that jamming attacks are less damaging in mmWave environments unless the attacker increases transmit power or reduces the distance to the receiver.

  \subsubsection{Wideband Robustness}  
  \label{sec:wideband}
  
  The most influential factor across all scenarios is \emph{channel bandwidth}. The improved performance of wider channels can be attributed to two main reasons:  
  
  \emph{(i) Higher achievable data rate due to increased bandwidth.}  
  Wider channels inherently support higher data rates at the same transmit power. 
  In our experiments, the offered traffic of 20~Mbps is well below the theoretical capacity even of the narrowest channels. 
  According to the Shannon–Hartley theorem~\cite{shanon}, channel capacity scales with bandwidth for a fixed \ac{snr}, so wider channels can carry more bits per second without increasing transmit power. 
  Operating far from saturation, the increased channel bandwidth thus provides additional margin against interference, allowing the system to maintain throughput and low latency under jamming.
  
  \emph{(ii) Reduced jammer Power Spectral Density.}  
  PSD measures how transmit power is distributed over a channel~\cite{psd}. For a transmitter with total power $P$ over bandwidth $B$, the average PSD is $\text{PSD (dBm/MHz)} = P_{\text{dBm}} - 10 \log_{10} B_{\text{MHz}}$. 
  Spreading the same jammer power over a wider bandwidth reduces interference per MHz. 
  For example, a 20~dBm jammer over 20~MHz yields a PSD of about 7~dBm/MHz. 
  Spreading the same power over 50~MHz reduces the PSD to roughly 3~dBm/MHz, lowering interference density. 
  This explains why wider channels experience less packet loss and higher throughput under identical jamming conditions.
  Combined, these two effects—higher inherent channel capacity and lower PSD of the jammer—work together to allow wideband FR2 channels to maintain better  performance under jamming conditions.
  In the next section, we translate our findings into practical guidance for deploying jamming-resilient 5G networks.
  
  \section{Practical Implications}
  \label{sec:implecations}
  As 5G becomes integral to mission-critical and industrial environments, ensuring network availability is essential~\cite{michaelides2025industry5G}. 
  For localized deployments, we recommend utilizing FR2 with wide channel bandwidths within a secured physical exclusion zone. 
  The high path loss of FR2, combined with restricted physical access, forces attackers to use significantly higher transmission power to achieve disruption~(cf.\ \S\ref{sec:takeaway}). 
  Our experiments show that wideband FR2 consistently maintains latencies below 10~ms—even under active jamming—meeting strict industrial requirements~\cite{michaelides2025industry5G}. 
  We also suggest adopting a high \ac{scs} as its impact on jamming resilience is negligible, yet it remains vital for low-latency performance~\cite{5Glatency}. 
  For larger-area deployments, FR1 with high channel bandwidth may offer a better trade-off between range and resilience. 
  Where both coverage and low latency are required, FR2 with massive MIMO and beamforming can mitigate path loss by focusing signals toward the UEs, effectively increasing the \ac{snr}~\cite{birutis2022study5G}.
  
\section{Future Work}

Future work can build on our open-source jammer to study application-specific deployments, such as V2X scenarios, where mobility/dynamic-topologies introduce additional challenges. 
  Moreover, investigating adaptive countermeasures, e.g., dynamic bandwidth allocation and beamforming strategies would provide deeper insight into practical defense strategies against jamming. 
  Finally, while our jammer is implemented using verified and calibrated classes (cf.\S\ref{sec:simulation}), providing a high degree of confidence, experimental validation in real-world deployments would further strengthen trust in the results.

\section{Conclusion}
  
In this work, we investigate the robustness of 5G’s flexible physical layer against jamming, motivated by its increasing integration into critical infrastructure and industrial deployments where availability is paramount. 
To this end, we develop and open-source an NR jammer in the ns-3 simulator and use it to evaluate how key 5G PHY features influence susceptibility to interference. Our results show that characteristics such as channel bandwidth and operating frequency significantly improve performance under jamming, while others, such as \ac{scs}, have negligible impact.
By releasing our implementation, we provide the means for operators and industrial stakeholders to model their own deployments and quantitatively assess their resilience against jamming under different configurations, supporting security-driven configuration decisions.

\section*{Acknowledgment}
Funded by the German Federal Ministry of Research, Technology and Space (BMFTR) under funding reference number 16KIS2409K (6GEM+) and the German Federal Office for Information Security (BSI) under funding reference number 01MO24003B (CSII). The authors are responsible for the content of this publication.

\end{document}

%% file: acronyms.tex
\DeclareAcronym{urllc}{
    short = URLLC,
    long  =  Ultra-Reliable Low Latency Communication
}
\DeclareAcronym{threegpp}{short=3GPP, long= 3rd Generation Partnership}

\DeclareAcronym{mmtc}{short= mMTC, long=Massive Machine-Type Communication}

\DeclareAcronym{sbi}{short= SBI, long=Service Based Interface}

\DeclareAcronym{ipsec}{short=IPsec, long= Internet Protocol Security}

\DeclareAcronym{up}{short =UP, long=User Plane}

\DeclareAcronym{cp}{short =CP, long=Control Plane}

\DeclareAcronym{ran}{short =RAN, long=Radio Access Network}

\DeclareAcronym{5gc}{short=5GC, long = 5G Core}

\DeclareAcronym{ue}{short=UE,long=User Equipment}

\DeclareAcronym{nf}{short=NF,long=Network Function}
\DeclareAcronym{gnb}{short=gNB, long=Next Generation Node B}

\DeclareAcronym{upf}{short=UPF,long=User Plane Function}

\DeclareAcronym{amf}{short=AMF,long=Authentication and Mobility Function}

\DeclareAcronym{smf}{short=SMF,long=Session Mangement Function}

\DeclareAcronym{pcf}{short=PCF, long = Policy Control Function}

\DeclareAcronym{lte}{short=LTE, long = Long Term Evolution}

\DeclareAcronym{tls}{short=TLS, long = Transport Layer Security}

\DeclareAcronym{ike}{short=IKEv2, long = Internet Key Exchange Version 2}

\DeclareAcronym{esp}{short=ESP,long = Encapsulation Security Payload }

\DeclareAcronym{psk}{short=PSK, long = Pre-Shared Keys}

\DeclareAcronym{nrf}{short=NRF, long = Network Repository Function}

\DeclareAcronym{supi}{short=SUPI, long= Subscription Permanent Identifier}

\DeclareAcronym{sim}{short =SIM, long = Subscriber Identity Module}

\DeclareAcronym{qos}{short=QoS, long = Quality of Service}

\DeclareAcronym{bbu}{short=BBU, long = Baseband Unit}

\DeclareAcronym{rrh}{short=RRH, long= Remote Radio Head}

\DeclareAcronym{cu}{short=CU, long= Control Unit}

\DeclareAcronym{ru}{short=RU, long= Radio Unit}

\DeclareAcronym{du}{short=DU, long= Distributed Unit}

\DeclareAcronym{cuup}{short=CU-UP, long= Control Unit-User Plane}

\DeclareAcronym{cucp}{short=CU-CP, long= Control Unit-Control Plane}

\DeclareAcronym{nas}{short=NAS , long= Non-Access Stratum}

\DeclareAcronym{rrc}{short=RRC, long = Radio Resource Control}

\DeclareAcronym{cpri}{short=CPRI, long =  Common Public Radio Interface}

\DeclareAcronym{ecpri}{short=eCPRI, long =  enhanced CPRI}

\DeclareAcronym{mno}{short=MNO, long= Mobile Network Operator}

\DeclareAcronym{aead}{short=AEAD, long=Authenticated Encryption with Associated Data }

\DeclareAcronym{dos}{short=DoS, long=Denial of Service }

\DeclareAcronym{scs}{short=SCS, long= subcarrier spacing }

\DeclareAcronym{psd}{short=PSD, long= power spectral density }

\DeclareAcronym{snr}{short=SINR, long= signal-to-interference-plus-noise ratio
}